# A framework for solving atomistic phonon-structure scattering problems in the frequency domain using Perfectly Matched Layer boundaries

Rohit R. Kakodkar[1] and Joseph P. Feser,[1,*]

[1]*Department of Mechanical Engineering, University of Delaware, Newark, DE 19716, USA*

*Author to whom correspondence should be addressed. Electronic mail: jpfeser@udel.edu*

Abstract: We present a numerical approach to the solution of elastic phonon-interface and phonon-nanostructure scattering problems based on a frequency-domain decomposition of the atomistic equations of motion and the use of perfectly matched layer (PML) boundaries. Unlike molecular dynamic (MD) wavepacket analysis, the current approach provides the ability to simulate scattering from individual phonon modes, including wavevectors in highly dispersive regimes. Like the Atomistic Green's Function (AGF) method, the technique reduces scattering problems to a system of linear algebraic equations via a sparse, tightly banded matrix regardless of dimensionality. However, the use of PML boundaries enables rapid absorption of scattered wave energies at the boundaries, and provides a simple and inexpensive interpretation of the scattered phonon energy flux calculated from the energy dissipation rate in the PML. The accuracy of the method is demonstrated on connected monoatomic chains, for which an analytic solution is known. The parameters defining the PML are found to affect the performance and guidelines for selecting optimal parameters are given. The method is used to study the energy transmission coefficient for connected diatomic chains over all available wavevectors for both optical and longitudinal phonons; it is found that when there is discontinuity between sublattices, even connected chains of equivalent acoustic impedence have near-zero transmission coefficient for short wavelengths. The phonon scattering cross section of an embedded nanocylinder is calculated in 2D for a wide range of frequencies to demonstrate the extention of the method to high dimensions. The calculations match continuum theory for long-wavelength phonons and large cylinder radii, but otherwise show complex physics associated with discreteness of the lattice. Examples include Mie oscillations which terminate when incident phonon frequencies exceeds the maximum available frequency in the embedded nanocylinder, and scattering efficiencies larger than two near the Brillouin zone edge.

## I. MOTIVATION

Understanding the propagation and scattering of vibrational waves with nanometer wavelengths (i.e. phonons) across heterogeneous interfaces is technologically imperative due to its relationship with thermal transport properties. For example, nanostructures embedded in semiconducting alloys are an effective method to scatter phonons and to subsequently reduce thermal conductivity in electronically active materials[1,2]. The dominant phonon wavelengths contributing to thermal conduction in alloys are usually a few nanometers long,[3] which is only slightly larger than interatomic distances. Another consequence of phonon interaction with interfaces is the observation that a thermal boundary resistance (TBR) exists even at the interface between molecularly conformal and smooth dielectric materials, caused when incident phonons experience reflection/scattering at the interface[4-8]. The magnitude of TBR can be related to the phonon dispersion and the energy transmission coefficient for phonons incident on the interface[7]. The dominant wavelengths contributing to TBR are short, typically falling near the Brillouin zone edge [9]. While well-accepted techniques exist for calculating and experimentally measuring phonon dispersions, methods for accurately calculating the energy transmission coefficient remain controversial; this is both because (1) the nature of phonon coherence across boundaries is controversial, (2) because the necessity of including accurate short wavelength modeling[10], and (3) inelastic collisions are thought be important in some situations such as at high temperature and interfaces of highly dissimilar materials[11].

Many techniques have been proposed to model phonon transport across heterogeneous interfaces. Initial attempts to explain TBR at low temperature utilized the acoustic mismatch model (AMM), in which isotropic continuum theory calculates the energy transmission coefficient from the acoustic impedance of the two materials being interfaced[4]. This model was found to be in poor agreement with experiments for certain types of interfaces such as helium-solid interfaces[7]. To account for higher frequency phonons, the diffuse mismatch model (DMM)[7,10], scattering mediated acoustic mismatch model (SMAMM)[6], lattice-dynamics models[5,8,12], and a number of atomistic computational techniques have been developed[13]. Lattice dynamics is ideally suited for modeling coherent transport in systems with homogeneous and periodic boundaries such as interfaces between smooth, epitaxial interfaces and in superlattices. The DMM has also been heavily utilized in literature as the DMM conveniently addresses two issues simultaneously: (1) it treats phonon interaction with the interface as incoherent elastic scattering, which has been argued to agree with experiment better than coherent theories[14]. (2) it allows transmission coefficients to be calculated at arbitrarily large wavenumber using only the phonon dispersion relations of the two materials being interfaced[10]. On the other hand, recent experimental observations indicate that for smooth and epitaxial films, coherent effects may be important[15-17].

Amongst coherent models for phonon transmission at smooth interfaces, wavepacket analysis within Molecular Dynamics (MD) is one highly developed spectrally-resolved approach[13,18]. MD is a time-domain approach in which a scattering system is modeled at an atomic level of detail and the interatomic interactions are based on potential functions. Wavepacket MD has been used both to study phonon transmission across single interfaces and to study scattering of phonons from embedded nanostructures. The MD approach performs simulations by preparing an isolated phonon wavepacket and observing its time evolution as it passes through an interface or scatterer[13,18]. MD simulations use the velocities and displacements of the atoms after the interaction has completed to determine properties such as reflectivity from an interface[13] or scattering cross section from isolated embedded nanostructures[18].

The ability of MD to model complex systems with atomistic detail is a major advantage over other available techniques. However, there are some important limitations. Since wavepackets must generally be initially located inside a homogenous region, a buffer region of atoms must be allocated that encompasses the entire wavepacket; however, if the extent of the wavepacket is made to be small to decrease computational requirements, then the spectrum of wavelengths represented in the wavepacket becomes broadened. Thus, MD models of phonon scattering do not track a single wavelength, and the breath of the spectrum has been as wide as 50% of the center wavelength for practical implementations[18]. In addition, small group velocity modes present computational challenges, since atoms may vibrate quickly, necessitating short time step, meanwhile the wavepacket itself moves very slowly. Thus, many timesteps are needed to track the wave as it passes through and beyond the scattering zone. Such dispersion may also spread the spatial extent of the wavepacket as it moves, which in turn may necessitate a larger buffer region before and after the scattering region that is required for the initial wavepacket. Non-linear interaction potentials (such as the Lennard-Jones potential) also require special attention be given to phonon-phonon interaction, which may be undesirable physics if the purpose of the simulation is to capture phonon-structure interactions.

The atomistic Green's function (AGF) method was developed to alleviate many of the issues surrounding MD wavepacket methods. Briefly, the method performs harmonic decomposition of the equations of motions for one or more isolated "contact" regions and a "device" region that contains the interface or scatterer, and then seeks to solve a self-consistent coupled problem using an associated connection matrix. The mathematics behind the method is highly developed, but ultimately reduces scattering problems to a linear set of equations via sparse/banded matrix for which atomic displacements in both the device region and contact regions are obtained. A recent review of the method and a summary of its applications was given by Sadasivam et al [19]. The method was originally developed to calculate phonon transmission coefficients[20], though a report has demonstrated extension of the method to calculate scattering rate of 3D embedded nanoparticles[21]. The method has also been integrated with Ab-Initio calculations to model realistic local bonding[22], and in one case has been amended using a Keldysh diagrammatic approach to include anharmonic effects[23].

In this work we present an alternative numerical route to the solution of elastic phonon scattering problems based on a frequency-domain formulation. The approach has similar advantages to AGF when compared to time-domain MD formulations, including the ability to simulate scattering from individual phonon modes (as opposed to wavepackets), wave-vectors in highly dispersive regimes including those with near-zero group velocity. The approach also reduces scattering problems to a system of linear algebraic equations via a sparse, tightly banded matrix. However, in contrast to AGF, the new formulation suggests a natural method of reducing the number of atoms needed in the "contact" region, as well as simple interpretation of the scattered phonon energy flux.

The approach is to approximate atomic interactions from the outset as harmonic and consequently ensure that interactions between phonons are elastic. In analogy to most analytical formulations of scattering problems, we consider an incident plane wave of known frequency/wavevector impinging with a scattering system and producing a scattered wave; by the harmonic approximation, all interactions must be at the same frequency, and thus the time dependence can be safely removed from the problem by assuming all displacement to be temporally sinusoidal and at the same frequency as the incident wave. The difficulty added by this approach is that having exact knowledge of the frequency requires incident and scattered waves that are in principle infinitely extended in space. In this work we address this difficulty by utilizing a technique ubiquitous in continuum computational formulations of electromagnetic and elastic scattering problems: Perfectly Matched Layers.[24]

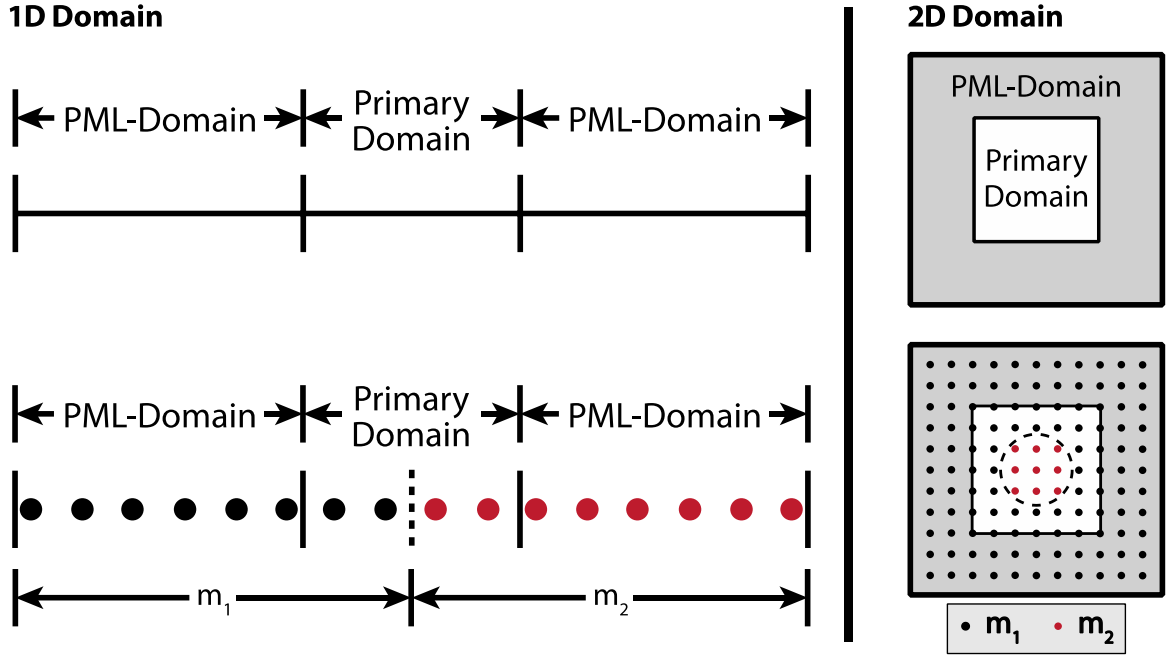

**FIG. 1.** Schematic diagrams of the phonon scattering simulation domains in one dimension (left) and two dimensions (right). Top diagrams show the general domain structure; bottom diagrams show examples of atomic placements inside the sub-domains.

Perfectly Matched Layers (PML's) are finite size domains, added to the edges of a simulation box, that are designed to absorb waves in a manner that minimizes reflections through complex impedance matching; thus, PML's are meant to computationally simulate infinitely long domains for wave scattering problems. PML's were originally developed as an aid to the solution of finite-difference time-domain electromagnetic wave scattering problems[24]. Subsequently, PML's have also been successfully applied to solve electromagnetic wave scattering problems in the frequency domain[25]. The attenuation of incident harmonic waves in PML's is a result of its analytic continuation in the complex plane[26]. Through this abstraction PML's have been applied to simulate infinite mediums for a wide variety of hyperbolic partial differential equations including electromagnetic, acoustic, and elastic wave scattering problems. Its application to intrinsically atomistic systems has been comparatively limited, though applications to time-domain MD simulations have been demonstrated. We formulate a frequency-domain PML for absorbing scattered waves by adapting an implementation originally developed for time-domain MD simulations[27,28].

The implementation and computational characteristics will be demonstrated in one-dimension and compared to exact analytical solutions available for connected semi-infinite monoatomic and diatomic chains over the entire Brillouin zone. In particular, we focus on the ability of the PML to absorb incident phonons without introduction of spurious back-reflections, and find that this places minimum constraints on the number of atoms that must be present in the PML and restricts the range of absorption coefficient of the PML. We then show the extension of the technique to multiple dimensions by demonstrating the calculation of phonon scattering cross section for an atomistically resolved embedded nanocylinder in a 2D crystal lattice.

## II. 1-D Crystal Lattice Model Formulation

Simulations are divided into a "primary domain" containing the scattering problem (i.e. the interface or lattice inhomogeneity) and PML domains connected to the edges of the primary domain (Fig. 1).

In one dimension, the model consists of a harmonic chain that may have inhomogenously distributed masses and spring constants. The notational conventions used for the masses and springs are given in Fig. 2.

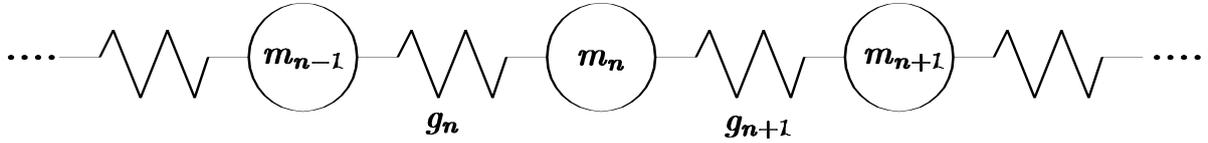

**FIG. 2.** Schematic of harmonic chain.

In the primary domain, applying Newton's law for nth atom yields

$$m_n \frac{d^2 u_n}{dt^2} = g_{n+1}(u_{n+1} - u_n) + g_n(u_{n-1} - u_n) \qquad 1$$

Where $g_n$ is the spring constant of the nth spring and $u_n$ is the displacement of the nth atom. Seeking solutions of the form $u_n = U_n e^{-i\omega t}$, Eq. becomes an algebraic equation in the frequency domain.

$$g_{n+1} U_{n+1} + (m_n \omega^2 - g_{n+1} - g_n) U_n + g_n U_{n-1} = 0 \qquad 2$$

Here, $U_n$ is the displacement of the nth atom and $\omega$ is the frequency of oscillation which is given by the dispersion relationship of the material. For scattering problems the total displacement can be considered as the superposition of an incident wave which satisfies the phonon dispersion of the primary medium and a perturbation caused by inhomogeneities in the system (i.e. a scattered wave), so $U_n = U_n^{inc} + U_n^{scat}$. In one-dimension, the incident wave frequency and displacements are often analytically calculable, but in general they can be computed as the eigenvalues/vectors of a crystals dynamic matrix for the homogenous phase. With the incident wave viewed as a known, Eq. 2 can be solved for the scattered wave amplitudes.

$$(-m_n\omega^2 + g_n + g_{n+1})U_n^{scat} - (g_{n+1}U_{n+1}^{scat} + g_n U_{n-1}^{scat})$$
$$= -(-m_n\omega^2 + g_n + g_{n+1})U_n^{inc} + (g_{n+1}U_{n+1}^{inc} + g_n U_{n-1}^{inc}) \qquad 3$$

Writing the Eq. 3 in a more suggestive form by separating the incident and the scattering wave terms we have, $AU = C$, where A is a tridiagonal matrix, C is a known vector of incident wave terms, and U is the unknown vector of scattered wave terms.

Numerically, the domain cannot continue to infinity so a PML is employed to absorb the scattered waves, which by energy conservation must be allowed carry energy out of the primary domain. Amongst various options for applying absorbing boundary conditions, perfectly matched layers are a good choice because the reflection coefficient between the primary domain-PML interface approaches zero for large PML thickness. Thus, the effect of a properly designed PML is to absorb wave energy without back-reflection, which makes it indistinguishable from an infinite medium of matched phonon dispersion.

A description of the application of PMLs for discrete atomic systems has been given by Li et al[27]. In one-dimension, the primary result is that inside the PML the equation of motion for each atom is modified such that the net force acting on each atom is

$$F = g_{n+1}(u_{n+1} - u_n) + g_n(u_{n-1} - u_n) - m_n\sigma^2 u_n - 2m_n\sigma \frac{du_n}{dt} \qquad 4$$

The first two terms on the right hand side are the interatomic forces; the last two terms represent an onsite restoration force and damping coefficient unique to the PML. Here σ is the damping rate coefficient whose value can be considered to be zero inside the primary domain of interest but is non-zero in the PML and may have spatial dependence. It has been shown that a parabolic spatial profile of the damping function is more efficient at absorbing spurious reflections than a linear one[29]. Hence, we have chosen a damping function given by,

$$\sigma = \frac{\sigma_{max}(x-L_{PML})^2}{L_{PML}^2} \qquad 5$$

Here $\sigma_{max}$ is the maximum value of the damping coefficient, which occurs at the PML edge furthest from the primary domain-PML interface, and $L_{PML}$ is the thickness of the PML region and $x$ is the distance from the domain-PML interface. The absorbing conditions is only applied to displacements attributable to the scattered wave (i.e. the incident wave travels through the layer unimpeded). Since the goal of 1D simulations is generally to calculate the transmission coefficient of the primary domain, in 1D the incident wave is applied only in the left or right PML domain. Then, converting Eq. 4 into the frequency domain,

$$\begin{aligned}(-m_n\omega^2 + g_n + g_{n+1} + m_n\sigma^2 - 2im_n\omega\sigma)U_n^{scat} - (g_{n+1}U_{n+1}^{scat} + g_n U_{n-1}^{scat}) \\ = -(-m_n\omega^2 + g_n + g_{n+1})U_n^{inc} + (g_{n+1}U_{n+1}^{inc} + g_n U_{n-1}^{inc})\end{aligned} \qquad 6$$

Thus, the only difference between the equations governing the PML region (Eq. 6) and the primary domain (Eq. 3) are the damping/restoring forces acting on the central atom of each equation. The governing equation inside the PML remains a set of linear algebraic equations $\hat{A}U = C$, but the diagonal elements of $\hat{A}$ differ from that of $A$. However, the right hand side, $C$, is identical since the damping coefficients of the PML don't act on the incident wave. The solution is insensitive to boundary condition assumed at the outer edges of the PML. For simplicity and to maintain minimum bandwidth of the linear equation set, we use a hard wall boundary, which reflects any unattenuated scattered wave through the PML and effectively doubles the length over which attenuation occurs. The absorbed wave energies can be calculated using the time-averaged rate of energy dissipation in the PML, $Q_{PML}$, which in 1D is given by

$$Q_{PML} = \sum_{PML} m_n \sigma_n \|U_n^{scat}\|^2 \omega^2 \qquad 7$$

which by energy conservation must also be the energy flux carried from the primary domain by the scattered waves. Thus, Eq. 7 is a convenient means to calculate the total energy reflection and transmission coefficients. Local energy fluxes in the primary domain may also be calculated using the scattered wave displacements, though these are not typically needed to calculate the transmission coefficients of interfaces or scattering cross section in higher dimensions.

At this point, it is possible to identify the differences between the PML-based formulation and the AGF. According to Sadasivam et al (see Eq. 20 of Ref. [19]) the form of the final set of equations being solved for a 1D chain by AGF is in exactly the same form as Eq. 6, except that the AGF perturbation term denoted $[0^+]$ is not identical, yielding the following important insights: (1) the AGF does not contain the onsite restoring term associate with impedance matching of the PML. (2) the AGF effectively uses a constant damping function, $\sigma$, rather than being spatially ramped one. From this, it is can be observed that AGF can effectively be considered similar to our formulation, but using an "absorbing boundary condition" (ABC) rather than a PML. This has important implications that will be revisited in Section IV.

We demonstrate the characteristics of the method by using it to predict the phonon reflection and transmission coefficients for connected semi-infinite monatomic and diatomic chains and comparing the results against analytically derivable ones throughout the entire Brillouin zone. The simulation domains are shown schematically on the left side of Fig. 1. The simulation cell consists of three distinct domains: a primary domain sandwiched between two PML domains. The primary domain contains the scattering problem to be solved; for example, to study the interface transmission coefficient between two monoatomic chains, the primary domain would contain the two chains connected at the center of the domain (see lower left of Fig. 1).

As a monatomic test case, mass and spring parameters for the left-most chain are chosen to have mass, sound speed, and lattice spacing similar to silicon ($m_1$=28 AMU, g=44.9 N/m, a=0.272 nm), which is interfaced with a chain of equal spring constant and lattice parameter, but differing mass, $m_2$. The total number of atoms in the simulation is 10,000 with 4,500 in each PML and 1,000 within the primary domain; the maximum damping coefficient is chosen as $\sigma_{max} = 1.68 \times 10^{12} s^{-1}$. The logic behind these PML parameters will be discussed shortly. The transmitted and reflected wave energies are calculated using rate of energy absorbed by each PML according to Eq. 7. The results are given in Fig. 3, which shows the simulated phonon energy transmission coefficient, $T_E$, for the interface as a function of dimensionless wavenumber over the entire Brillouin zone for several values of the chain mass ratio, $r \equiv m_1 / m_2$. The numerical solutions visually coincide with the analytically known solution[5].

For the case $r < 1$, the maximum phonon frequency in the bulk of medium 1 is lower than in medium 2. Thus, over the entire range of incident phonon frequencies, there exists a propagating mode to transmit energy to in medium 2. For $k \to 0$, the acoustic mismatch model (AMM) result[7] is recovered; for wavevector comparable to the size of the Brillouin zone $T_E$ is suppressed and becomes zero at the edge of the Brillouin zone where no modes of propagation exist in material 1. In addition to lowering the transmission coefficient as $k \to 0$, smaller r has the effect of broadening the portion of the Brillouin zone that experiences sub-AMM transmission coefficients.

For the case $r > 1$, the maximum phonon frequency in material 2 is less than in material 1. Since anharmonicity is absent from the model, it is impossible for some high frequency phonons in material 1 to be transmitted into material 2. This is evident in Fig. 3 as transmission coefficients that go to identically zero when there are no available phonon modes in material 2. In this sense, using coherent models with full dispersion replicates one of the most important features of the diffuse mismatch model (DMM), namely that transmission coefficients drop when there are no shared modes, but without invoking assumptions about incoherent scattering/loss of phonon history.

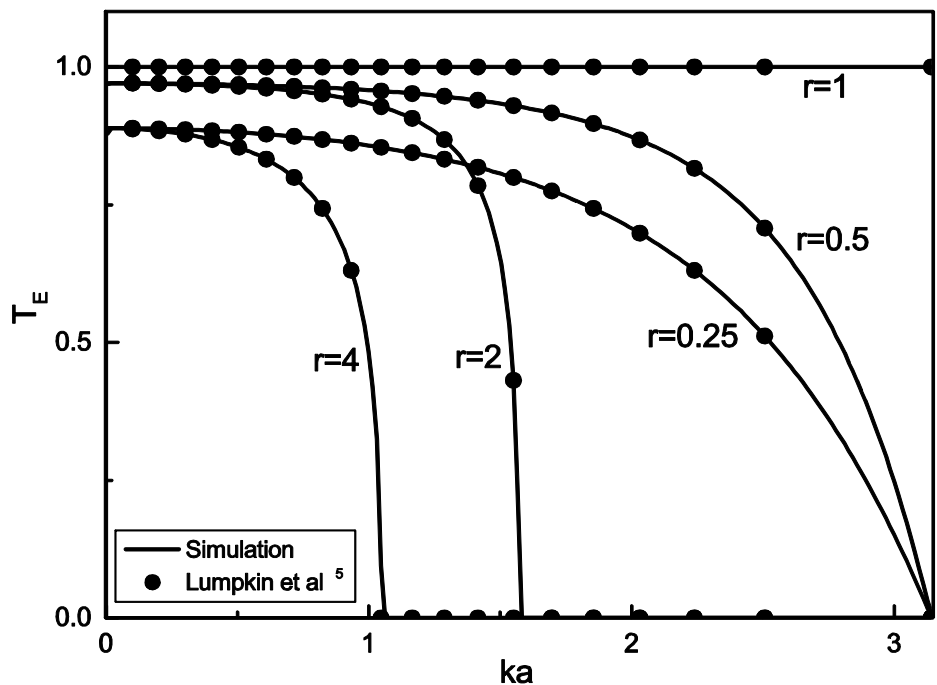

**FIG. 3.** Comparison of numerical and analytical results for transmission coefficient of a monatomic chain interface for several mass ratios.

### III. Choice of PML Parameters

One important aspect of implementing the scheme above is choosing an appropriate thickness, $L_{PML}$, and characteristic attenuation strength, $\sigma_{max}$, for the PML regions. In particular, each atom added to the PML adds an equation to be solved to the system of linear equations (Eq. 6), and is thus computationally expensive both in terms of speed and memory usage. Therefore, the thickness of the PML must be optimized to balance the desire for accuracy, which generally benefits from a longer PML, against the desire to minimize computational requirements. While some guidance exists in literature regarding how to choose shape of the spatial dependence of the damping function, we have not found any guidance regarding how to choose optimal PML length and characteristic strength. Therefore, the accuracy of the method is investigated here as a function of the PML design parameters for the case of connected monatomic chains (mass ratio, $r = 0.5$), where the analytical result for the infinitely extended system is known[5].

To characterize the accuracy we calculate the relative error, $\varepsilon \equiv |T_{exact} - T_{simulation}|/T_{exact}$, where T is the energy transmission coefficient from the interface between the connected monatomic chains for the exact and simulation results. We hypothesize that any errors that arise must be a result of non-zero reflection coefficient between the primary domain and the PML interfaces. The spurious reflection at each primary-domain/PML interface could at most be a dependent on $\sigma_{max}$, $L_{PML}$, $\lambda$, $a$, and $v_g$ where the last three parameters are the phonon wavelength, lattice constant, and group velocity respectively. By dimensional analysis, the number of independent variables can be reduced by two, so that the error is necessarily a function of only three dimensionless parameters, $\alpha_I$ ($I = 1, 2, 3$). One convenient choice to characterize the error is the set of parameters:

$$\alpha_1 \equiv \frac{L_{PML}}{\lambda} \qquad \qquad 8$$

$$\alpha_2 \equiv \frac{\sigma_{max} L_{PML}}{v_g} \qquad \qquad 9$$

$$\alpha_3 \equiv \frac{\lambda}{a} \qquad 10$$

These are chosen for their physical interpretation. $\alpha_1$ is the ratio of the length of the PML to the phonon wavelength. $\alpha_2$ is the ratio of the time a wavepacket would spend traversing the PML to the attenuation time. Intuitively, one expects that for sufficiently small $\alpha_2$, scattered/reflected waves will not be completely attenuated within the PML, which will manifest as errors in the computed reflection coefficient. $\alpha_3$ is the wavelength written in units of the lattice constant. Physically, it is anticipated that for long wavelength ($\alpha_3 \gg 2$) the system will behave as a continuum and the error, $\varepsilon = \varepsilon(\alpha_1, \alpha_2)$ only.

Figure 4 shows contour plots of the relative error on a logarithmic scale ($\log_{10} \varepsilon(\alpha_1, \alpha_2)$) at long wavelength ($\alpha_3 = 15$) and short wavelength ($\alpha_3 = 3$). We can make 3 general observations: (1) when $\alpha_2$ is sufficiently small, the PML is not thick enough to completely attenuate the scattered wave. This results in the scattered wave being backreflected into the primary domain via the hard wall boundary condition, and a subsequent error. To improve accuracy $\alpha_2$ can always be made arbitrarily high without affecting the computational speed by choosing an appropriate value of the PML attenuation parameter, $\sigma_{max}$. .(2) However, for values of $\alpha_2$ that are too high, a steep gradient in the damping coefficient can also cause spurious backreflections. Therefore, for any given length of PML ($\alpha_1$), there is an optimum damping coefficient ($\alpha_2$) for highest accuracy. (3) To obtain accuracy of several percent, each PML must be at least several wavelengths long. This sets a lower bound to the total size of the simulation box ($\alpha_1$) and is the primary factor that determines computational time and memory usage characteristics of the method. The accuracy of the simulation may be increased to by orders of magnitude by increasing the length of the PML at the expense of increased computation time and memory usage. Depending on these characteristics, there is an optimal set of PML parameters for a given desired level of accuracy. By taking a sufficiently large $\alpha_1$ and optimal $\alpha_2$, it is possible to reach machine precision accuracy ($\varepsilon$ ~$10^{-14}$ for double precision) using this method, though this is seldom necessary. Although we restrict the scope of results to one dimension here, this result should serve as a useful guideline for choosing PML parameters in multidimensional simulations as well, where tradeoffs between computation expense and accuracy are of greater concern.

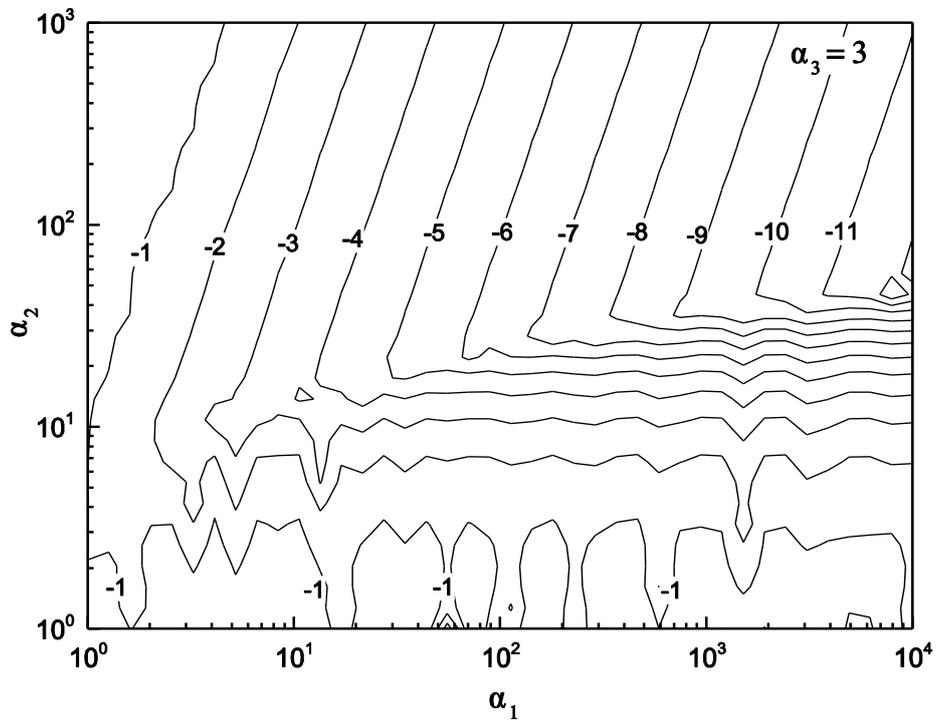

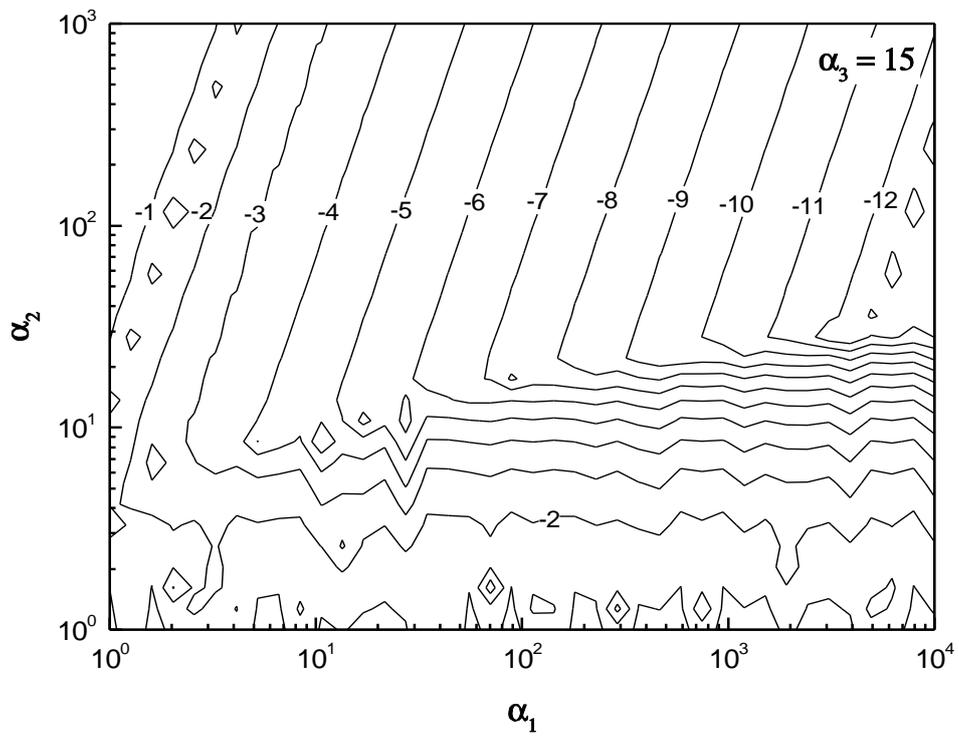

**FIG. 4.** Contour plot of the relative error in the energy transmission coefficient on a logarithmic scale, $\log_{10} \varepsilon(\alpha_1, \alpha_2)$, for wavelengths $\lambda = 3a$ and $\lambda = 15a$.

In one dimension, the model can also be used to study more complicated scenarios. Even applied to connections between diatomic chains, the model gives useful physical insights, as we will show now. To our knowledge, the analytic solution for the reflection coefficient at interfaces of diatomic chains has not been given previously in literature, but it is straightforward to derive the result and this is given in an Appendix. The addition of phonon band gaps and optical phonons increases the number of possible simulation scenarios such that we cannot provide an exhaustive list of observations. However, one particularly enlightening application is to analyze what happens if two diatomic chains with identical acoustic impedance, but with sublattices that are discontinuous, are interfaced. Examples of such interfaces occur in practical epitaxial systems such as type II heterojunctions of GaSb/InAs. In this case, each material being interfaced has identical mass ratio in their sublattice (i.e. $m_{Ga}/m_{Sb} \approx m_{As}/m_{In}$) and the bonding type and lattice parameter is nearly the same. However, the mass of cation and anion are reversed. These materials have nearly the same acoustic impedance and even the same phonon dispersion; thus the AMM would conclude that there is zero reflectivity from their interface. The diffuse-mismatch model (DMM), however, would predict that the reflection coefficient was ½ and independent of frequency. Neither of these assumptions would seem valid since their sublattice masses become discontinuous upon interfacing, yet there is no randomness near the surface to lead to incoherent scattering as would be implicitly assumed in the DMM. To understand the reflection behavior of this interface we have performed simulations, using the same PML parameters as in the monatomic case. The values of parameters used to approximately represent a GaSb/InAs heterojunction are $m_{Ga}$ = 72AMU, $m_{Sb}$ =117AMU, $m_{In}$ = 117, $m_{As}$ =72AMU g=26.14N/m and a=0.303nm.

The results are given in Fig. 5. In this case we separate transmission behavior for incident acoustic phonons from that of optical phonons. Note that the traditional AMM yields a prediction for the transmission coefficient of long-wavelength acoustic phonons, but gives no insight to the behavior of optical phonons because optical phonons do not exist in continuum mechanical formulations. The result of the simulation indicates that for acoustic phonons, the transmission coefficient is unity at the center of the Brillouin zone as predicted by the AMM, but goes to zero at the zone edge and is suppressed throughout much of the Brillouin zone.

For optical phonons, the transmission coefficient is generally low across the Brillouin zone reaching exactly zero at both the center and edge. Thus, even for epitaxial materials with similar acoustic impedance and identical phonon density of states, there is reason to expect low phonon transmission coefficient for short wavelength phonons (both optical and acoustic) that tend to dominate thermal phonon transport at interfaces. These results suggest that it is necessary to consider continuity of the sublattices when trying to design materials with high phonon transmission coefficient for thermal interface conductance applications. We note that one of the highest thermal interface conductances at ambient conditions that has been experimentally measured is for an epitaxial TiN/MgO interface[14], which shares both crystal structure (rock salt) and one similar sublattice (O/N) across the interface.

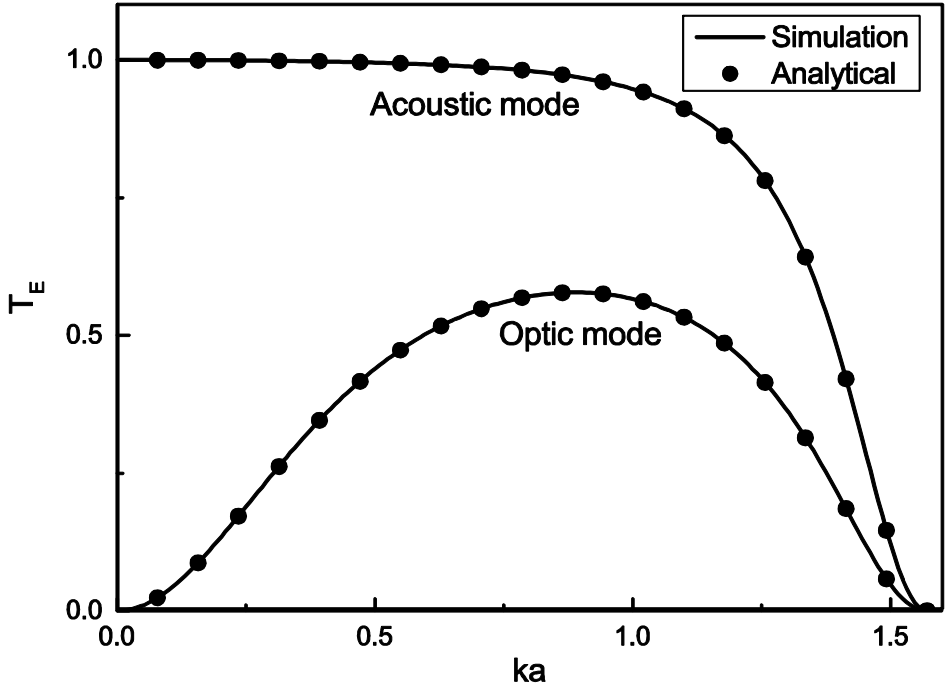

**FIG. 5.** Transmission coefficient of connected diatomic chains with GaSb/InAs-like mass ratios.

## IV. EXTENSION TO HIGHER DIMENSIONS

We now demonstrate the extension of the method to higher dimensions by calculating the scattering cross section of an embedded cylindrical impurity in a host material over a wide range of wavelengths ($2a < \lambda < 60a$ or $0.54$ nm $< \lambda < 16$ nm).

In higher dimensions, the PML fully encapsulates the primary domain, rather than forming two separate contacts (Fig. 1b). Also, since the scattered waves can have both longitudinal and transverse polarizations, it is necessary to provide PML damping/stiffness modifications in both x- and y- directions (and z in 3D). For simplicity, we chose a demonstration case where both the host material and cylindrical region are monoatomic simple cubic materials with the same lattice and elastic constants, but with different atomic masses. Nearest and second nearest neighbor interactions are considered (8 connections per atom, Fig. 6). The harmonic decomposition of the x- and y- momentum equations in this case are given by Eq. 11 & 12 respectively (where $\phi \equiv m_P \omega^2 - 2[g_1 + g_2]$ ), and are applied to each atom in the simulation except those on the PML outer edge where we use a zero displacement condition for the scattered wave displacements.

$$\begin{bmatrix} g_2/2 \\ g_2/2 \\ g_1 \\ g_2/2 \\ -g_2/2 \\ \left(\phi + i\left[2m_P\sigma_x\omega + im_P\sigma_x^2\right]\right) \\ g_2/2 \\ -g_2/2 \\ g_1 \\ g_2/2 \\ g_2/2 \end{bmatrix}^T \begin{bmatrix} U_{x,SW}^{scat} \\ U_{y,SW}^{scat} \\ U_{x,W}^{scat} \\ U_{x,NW}^{scat} \\ U_{y,NW}^{scat} \\ U_{x,P}^{scat} \\ U_{x,SE}^{scat} \\ U_{y,SE}^{scat} \\ U_{x,E}^{scat} \\ U_{x,NE}^{scat} \\ U_{y,NE}^{scat} \end{bmatrix} = - \begin{bmatrix} g_2/2 \\ g_2/2 \\ g_1 \\ g_2/2 \\ -g_2/2 \\ \phi \\ g_2/2 \\ -g_2/2 \\ g_1 \\ g_2/2 \\ g_2/2 \end{bmatrix} \begin{bmatrix} U_{x,SW}^{inc} \\ U_{y,SW}^{inc} \\ U_{x,W}^{inc} \\ U_{x,NW}^{inc} \\ U_{y,NW}^{inc} \\ U_{x,P}^{inc} \\ U_{x,SE}^{inc} \\ U_{y,SE}^{inc} \\ U_{x,E}^{inc} \\ U_{x,NE}^{inc} \\ U_{y,NE}^{inc} \end{bmatrix}^T \quad 11$$

and

$$\begin{bmatrix} g_2/2 \\ g_2/2 \\ -g_2/2 \\ g_2/2 \\ g_1 \\ \phi + i\left[2m_P\sigma_y\omega + im_P\sigma_y^2\right] \\ g_1 \\ -g_2/2 \\ g_2/2 \\ g_2/2 \\ g_2/2 \end{bmatrix}^T \begin{bmatrix} U_{x,SW}^{scat} \\ U_{y,SW}^{scat} \\ U_{x,NW}^{scat} \\ U_{y,NW}^{scat} \\ U_{y,S}^{scat} \\ U_{y,P}^{scat} \\ U_{y,N}^{scat} \\ U_{x,SE}^{scat} \\ U_{y,SE}^{scat} \\ U_{x,NE}^{scat} \\ U_{y,NE}^{scat} \end{bmatrix} = -\begin{bmatrix} g_2/2 \\ g_2/2 \\ -g_2/2 \\ g_2/2 \\ g_1 \\ \phi \\ g_1 \\ -g_2/2 \\ g_2/2 \\ g_2/2 \\ g_2/2 \end{bmatrix} \begin{bmatrix} U_{x,SW}^{inc} \\ U_{y,SW}^{inc} \\ U_{x,NW}^{inc} \\ U_{y,NW}^{inc} \\ U_{y,S}^{inc} \\ U_{y,P}^{inc} \\ U_{y,N}^{inc} \\ U_{x,SE}^{inc} \\ U_{y,SE}^{inc} \\ U_{x,NE}^{inc} \\ U_{y,NE}^{inc} \end{bmatrix}^T \qquad 12$$

The parameters in Table 1 simulated a silicon-like material with the atomic mass of the cylinder atoms $m_{cyl}/m = 4$. The ratio of spring constants is set to $g_2/g_1 = 1/2$ to create an isotropic material in the limit $\lambda \gg a$. This facilitates direct comparison with an exact continuum solution developed by White[30] for elastic wave scattering from a cylindrical discontinuity in an isotropic medium for arbitrary incident polarization and wavelength, for which the underlying assumptions are valid for $\lambda, D \gg a$.

| Parameter | Host Material | Scattering Material |
|---|---|---|
| Crystal Structure | Simple Cubic | Simple Cubic |
| lattice parameter, $a$ | 0.273 nm | 0.273 nm |
| atomic mass, $m$ | $1.735 \times 10^{-15}$ kg | $6.940 \times 10^{-15}$ kg |
| spring constant, $g_1$ | $15.92 \times 10^{12}$ N/m | $15.92 \times 10^{12}$ N/m |
| spring constant, $g_2$ | $7.96 \times 10^{12}$ N/m | $7.96 \times 10^{12}$ N/m |
| sound speed, long. $v_L$ (calc) | 10110 m/s | 5055 m/s |
| sound speed, trans. $v_T$ (calc) | 5840 m/s | 2920 m/s |

Table 1. Parameters used in scattering cross section calculation whose result are shown in Fig. 7.

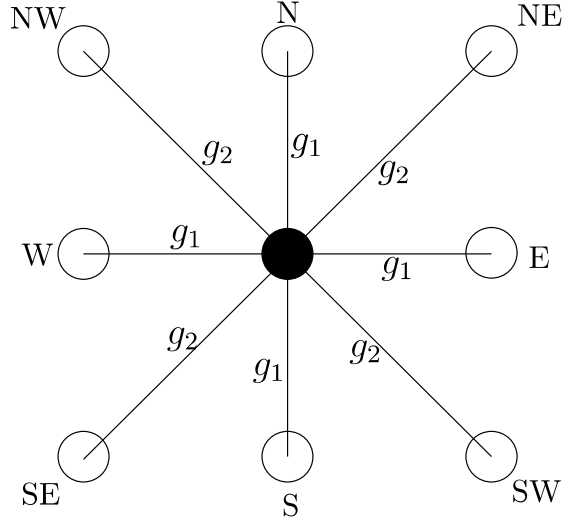

**FIG. 6.** 2D spring system around atom P

For all 2D simulations a computational domain size of 400 atoms × 400 atom is used with a primary domain of 100 atoms x 100 atoms. Thus, the PML is 150 atoms long on each side. According to Fig. 4, if we choose an appropriate damping coefficient, $\sigma_{max}$, the simulations will have relative accuracy to better than 10% for $\lambda \lesssim N_{PML} a / 2$ ($\alpha_1 \gtrsim 2$) and much better accuracy at shorter wavelengths for fixed PML thickness, such as those near the Brillouin zone edge. Thus, $\alpha_2 = 30$ is chosen (Fig. 4) and scattering cross section for incident waves in the range $2a < \lambda < 60a$ simulated for several different cylinder diameters, $D = \{8a, 16a, 32a\}$. The scattering cross section, $\gamma$ is defined as the time-averaged rate of energy being carried away by the scattered wave (Watts) normalized by the energy intensity of the incident wave (Watts/m in 2D). The former can be easily calculated in the current context by measuring the time-averaged total energy dissipation in the PML, which in 2D is

$$Q_{PML} = \sum_{PML} m_n \left( \sigma_{x,n} \left\| U_{x,n}^{scat} \right\|^2 + \sigma_{y,n} \left\| U_{y,n}^{scat} \right\|^2 \right) \omega^2 \qquad 13$$

The incident wave in this case is applied throughout the entire host material including the PML (everywhere except for the nanocylinder), although the same scattering cross section (dissipated energy in the PML) would be obtained if the incident wave had been applied only in the PML. The advantage of the former method is that by applying the incident wave throughout the entire host material, including the primary domain, the analogous scattered displacement field as obtained from analytical methods is obtained, rather than a superposition of the incident and scattered waves. The calculated scattering cross sections are given in Fig 7 for incident longitudinal and transverse waves traveling in the x-direction.

Results are presented in terms of scattering efficiency $\gamma/D$ of the cylindrical impurity and plotted as a function of the dimensionless scattering parameter ($kD$).

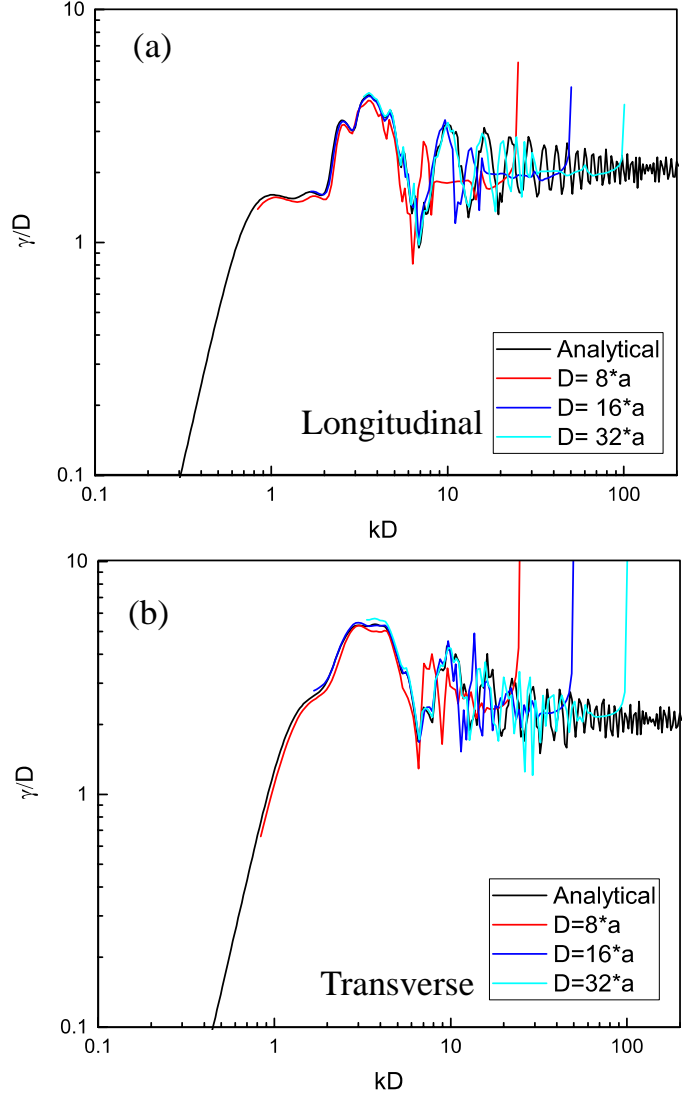

**FIG. 7.** Scattering efficiency of an embedded 2D cylinder due to (a) longitudinal incident phonons and (b) transverse incident phonons. Numerical calculations are compared to exact results from continuum theory[30].

Continuum theory predicts $\gamma/D \sim (kD)^3$ in the Rayleigh regime ($kD \ll 1$) and a scattering efficiency oscillating about $\gamma/D = 2$ in the geometric regime ($kD \gg 1$) with a series of Mie oscillations in between. Our atomistic results show agreement with continuum theory when $\lambda \gg a$, and also show some unique features of scattering that arise when atomistic resolution is included. Some effects that are visible from our simulations include the following:

a) Unlike continuum theory, a discrete lattice has a minimum physically significant wavelength (or equivalently maximum wavenumber) associated with the edge of the Brillouin zone. Thus, it is only physically meaningful to plot scattering cross section to a maximum value of $kD$.

b) Since the scattering material has a maximum phonon frequency that is smaller than the host material's in our test case, incident phonons exceeding this frequency can only couple to evanescent waves in the scatterer. This prevents waves from propagating through the cylinder that would otherwise cause the constructive and destructive interfering waves associated with Mie scattering. Thus Mie oscillations disappear beyond a critical incident wavenumber/frequency and the scattering cross sections settle to $\gamma/D \approx 2$ more quickly than predicted in continuum theory.

c) Although the simulated material is designed to have isotropic sound speed, the material is still anisotropic for mid- and short-wavelength phonons, and also shows significant phonon dispersion leading to zero group velocities at the Brillouin zone edge. We find that that scattering efficiency rises above 2 when wavelengths approach the zone edge and diverges near the zone edge. For scattering of longitudinal waves, there is a visible effect that occurs because part of the scattered wave is carried away by transverse scattered modes. Note that just as transmission coefficients are definable for phonons near the zone edge with dispersion, the scattering cross section, as we have defined it, is still physical meaningful and calculable near the zone edge.

d) Due to the discrete nature of the lattice, the scatterer shape cannot exactly replicate a cylinder for small cylinder diameters. This becomes an issue when the scatterer size is comparable to that of the lattice dimension; this can be seen in Fig 7 where the graph for $D=32a$ has better agreement with the continuum results compared to that of $D=8a$. In the $D=8a$ case, the total mass of the discrete scatterer is 23% lower than the total mass that would be associated with a continuum cylinder of the same diameter and bulk density.

In general, the computational requirements are significantly more demanding as the number of dimensions, $n$, is increased. The total number of linear equations to be solved is equal to the number of atoms in the simulation box, including those in the PML regions. The number of equations, $N_{eq}$, to be solved for an n dimensional system with symmetric simulation box is $N_{eq} = nN^n$, where $N$ is the number of atoms along a single dimension, and the bandwidth of the defining matrix scales as $BW \sim N^{n-1}$; thus the defining matrix is not tridiagonal in higher dimensions, although $BW \ll N_{eq}$. Note that the sparsity pattern of the matrix depends on the lattice structure and on the cut-off distance, $d$, for atomic interactions. Increasing the cut-off distance increases both the bandwidth ($\sim d$), which increases the solver time ($\sim d^2$, for direct solvers) and the number of non-zero elements of the matrix ($\sim d^n$), which leads to rapidly increasing memory requirements. The minimum feasible size of the simulation box, $N$, depends on the maximum wavelength to be simulated, since each PML must be several wavelengths thick (Fig. 4). Hence in higher dimensions, it is critical to use a PML with the least number of atoms possible. For example, in the 2D scattering cross section calculation above, 94% of the atoms are located in the PML so the number of atoms in the PML is the primary factor that determines the computational memory requirements as well as the time required to solve the system of equations. As such this manuscript provides guidance in how to choose optimal PML parameters (Fig. 4). The current approach has proven successful for total simulation boxes up to 2.56 million atoms in 2D on a shared memory computer system with 192 GB of RAM using a direct solver; using an iterative solver should improve this in the future.

Note that numerical memory requirements are not dissimilar to MD, which would track $2N_{eq}$ variables *at each time step*. In contrast, the current approach solves a linear system of $N_{eq}$ equations only once for unknown displacements $U_n^{scat}$, which is aided by the fact that the defining matrix is sparse, tightly banded ($BW \sim O(N^{n-1})$), and diagonally dominant. The memory wasted simulating atoms in the PML region is also comparable to MD since MD wavepackets need several wavelengths of space between their initial location and the scattering interface. While the memory requirements are similar to MD, there are other advantages of using frequency-domain approach. Since the simulation time is independent of the group velocity it is computationally simpler to obtain scattering properties near the edge of the Brillouin zone and for materials with optical phonon modes. In addition, by eliminating the need for wavepacket simulation, the method has the ability to obtain scattering results for individual phonon modes since the incident wave is infinitely extended in space. On the other hand, employing MD on distributed memory parallel computers is more straightforward than the current approach, which maybe a significant advantage for large problems.

A comparison to AGF is also instructive and may point toward other uses of the PML approach. The accuracy of our method is centered on the ability of the PML to effectively attenuate the scattered waves at the simulation boundaries without spurious reflections. Since PML's are impedence matched with the primary domain material there is minimal reflection from primary domain-PML boundary optimal chosen $\sigma_{max}$ (Fig. 4); this is not the case for the contacts used in AGF, which can be viewed as simple absorbing boundary conditions (ABC) as discussed in Section II. It is has been shown that PML's are more efficient for attenuating waves without reflection than simple ABC's[31]. AGF implementations cleverly avoid the need to simulataneously solve for the discplacements in the contacts using a decimation technique[19], thereby freeing memory associated with the contact atoms in exchange for having to invert matrices of smaller size, $N_C \times N_C$, many times during course of the decimation iterations, where $N_C$ is the number of atoms along the contacts. In the AGF method, the speed of convegence of the decimation technique and the accuracy depend on the strength of the ABC ($[0^+]$); this suggests that the damping term $[0^+]$ appearing in the contact matrix (see Eq. 20 of Ref. [19]) could be modified by using terms corresponding to a PML to achieve faster convergence and higher accuracy for the AGF method.

The formulation does present an obvious limitation in that inelastic scattering cannot be included due to the assumption of harmonic bonding. This is problematic for capturing the physics of highly mismatched systems such as Pb/diamond[11], where experimental evidence supports energy transfer mediated by inelastic scattering. The inclusion of anharmonic effects is possible through perturbation theory, but is computationally expensive. However, the lack of anharmonic physics may also be an advantage in situations where it is desirable to separate phonon-phonon interactions from phonon-structure interactions.

## V. CONCLUSIONS

A framework for numerically solving elastic phonon scattering problems has been described and demonstrated in one and two dimensions. The method uses a frequency-domain decomposition of the harmonically-approximated equations of motion and perfectly matched layer boundaries to reduce scattering problems to a set of linear algebraic equations with low matrix bandwidth. The approach accurately reproduces available analytic results. By tracking the energy dissipation in the PML regions, the method provides a straightforward quantitative tool for measuring the scattered wave energy.

Even applied in one-dimension, the model gives physical insight into a number of fundamental interface phenomena: For a monoatomic interface with dissimilar atomic masses, the transmission coefficients for short wavelength phonons are found to be lower than the continuum AMM prediction and identically zero when no modes of equal energy are available for transmittance. For interfaces between diatomic chains, it is found that when there is discontinuity between sublattices, even connected chains of equivalent acoustic impedence have near-zero transmission coefficient for short wavelengths acoustic phonons. For optic phonons in the same material, the transmission coefficients are also non-unity, reaching zero at both the Brillouin zone edge and center. The observations point to the need to maintain sublattice continuity in order to increase transmission coefficient

The extension of the method to multiple dimensions is demonstrated through the calculation of the scattering cross section of a cylindrical embedded scatterer over a wide range of wavelengths. The simulation results are in agreement with continuum theory for $\lambda, D \gg a$, where continuum theory is valid. Unique scattering features due to lattice discreteness are observed such as cessation of Mie Oscillations when coupled modes are not present in scatterer and phonon dispersion effects.

Acknowledgements: We gratefully acknowledge funding from the University of Delaware Research Foundation.

**Appendix: Phonon Reflection Coefficient for an Interface of Diatomic Chains**

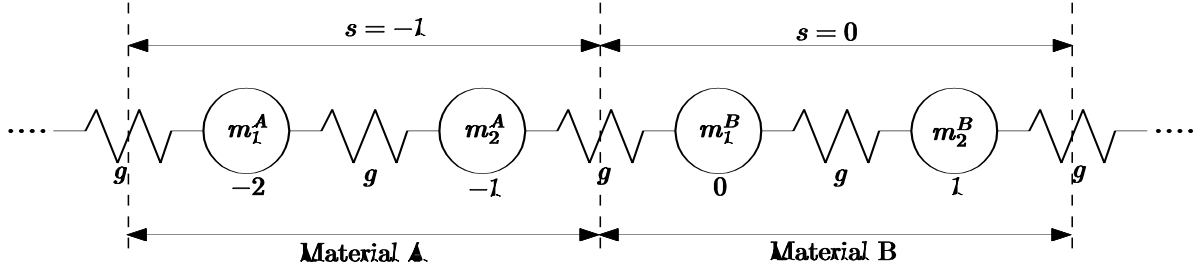

**FIG. 8.** Schematic diagram of diatomic interface

The reflection and transmission coefficient for two connected diatomic crystal chains (material A and B) can be calculated as follows, referring to Fig. 8 for notational conventions. Assume the displacements of incident, reflected and transmitted waves to be of the form given below

Incident wave :
$$u^I_{2s+1} = u^I_0 \xi^I e^{i[(2s+1)k_1(\omega)a - \omega t]}$$
$$u^I_{2s} = u^I_0 \eta^I e^{i[(2s)k_1(\omega)a - \omega t]}$$

Reflected wave:
$$u^R_{2s+1} = u^R_0 \xi^I e^{i[-(2s+1)k_1(\omega)a - \omega t]}$$
$$u^R_{2s} = u^R_0 \eta^I e^{i[-(2s)k_1(\omega)a - \omega t]}$$

Transmitted Wave:
$$u^T_{2s+1} = u^T_0 \xi^T e^{i[(2s+1)k_2(\omega)a - \omega t]}$$
$$u^T_{2s} = u^T_0 \eta^T e^{i[(2s)k_2(\omega)a - \omega t]}$$

where $u_0$ represents the amplitude of each wave. $\xi$ and $\eta$ are components of the unit eigenvector obtained by solving the eigenvalue problem for the diatomic crystal matrix[32],

$$\begin{bmatrix} -\dfrac{2g}{m_1} & \dfrac{2g\cos(ka)}{m_1} \\ \dfrac{2g\cos(ka)}{m_2} & -\dfrac{2g}{m_2} \end{bmatrix} \begin{bmatrix} \xi \\ \eta \end{bmatrix} = -\omega^2 \begin{bmatrix} \xi \\ \eta \end{bmatrix} \quad 14$$

There are two sets of solutions representing the acoustic and optical modes. The frequency, polarization, and amplitude of the incident wave are considered known, which allows for the unique solution of wavenumber ($k_1(\omega)$), $\xi^I$ and $\eta^I$. Knowing $\omega$ allows the transmitted wave parameters $\xi^T$, $\eta^T$ as well as the appropriate wavevector ($k_2(\omega)$) to be determined. Note that the wavevectors of the reflected and transmitted waves must be chosen to have a group velocity that carries energy away from the interface if the incident wave is chosen so as to carry energy toward the interface. The amplitude of the reflected and transmitted waves are the only remaining unknowns. These are found by applying Newton law at each of the two interfacial atoms, which yields

$$\begin{bmatrix} \dfrac{m_2^A \omega^2}{g}\eta^I - 2\eta^I + \xi^I e^{ik_1(\omega)a} & \xi^T e^{ik_2(\omega)a} \\ \eta^I & e^{ik_2(\omega)a}\left(\dfrac{m_1^B \omega^2}{g}\xi^T - 2\xi^T + \eta^T e^{ik_2(\omega)a}\right) \end{bmatrix} \begin{bmatrix} u_0^R / u_0^I \\ u_0^T / u_0^I \end{bmatrix} = \begin{bmatrix} -\dfrac{m_2^A \omega^2}{g}\eta^I + 2\eta^I - \xi^I e^{-ik_1(\omega)a} \\ -\eta^I \end{bmatrix}$$

Which can be solved to obtain the amplitude of the displacements. To calculate the energy transmission coefficient, the energy reflection coefficient can be first calculated as $R_E = (u_0^R / u_0^I)^2$, from which the energy transmission coefficient can be calculated as $T_E = 1 - R_E$ by conservation of energy.